\documentclass[runningheads]{llncs}

\usepackage{cite}
\usepackage{amsmath,amssymb,amsfonts}
\usepackage{algorithmic}
\usepackage{graphicx}
\usepackage{textcomp}
\usepackage{xcolor}
\usepackage{comment}

\usepackage{nth}
\usepackage{tikz}
\usepackage{pgfplots}
\usepackage{subcaption}
\usepackage{booktabs}
\usepackage{siunitx}
\usepackage{multirow}
\usepackage{makecell}
\usepackage{cleveref}
\usepackage{csquotes}

\usetikzlibrary{external}
\usepackage{abbr}
\usepackage{style}

\begin{document}

\title{$\text{A}^3$: Activation Anomaly Analysis}


\author{Philip Sperl\thanks{Philip Sperl and Jan-Philipp Schulze are co-first authors.} \and
Jan-Philipp Schulze\inst{\star} \and
Konstantin B\"ottinger}
\authorrunning{Philip Sperl, Jan-Philipp Schulze, and Konstantin B\"ottinger}
\institute{Fraunhofer Institute for Applied and Integrated Security \\
\email{\{philip.sperl, jan-philipp.schulze, konstantin.boettinger\}@aisec.fraunhofer.de}
}

\maketitle

\begin{abstract}
Inspired by recent advances in coverage-guided analysis of neural networks, we propose a novel anomaly detection method.
We show that the hidden activation values contain information useful to distinguish between normal and anomalous samples.
Our approach combines three neural networks in a purely data-driven end-to-end model.
Based on the activation values in the target network, the alarm network decides if the given sample is normal.
Thanks to the anomaly network, our method even works in strict semi-supervised settings.
Strong anomaly detection results are achieved on common data sets surpassing current baseline methods.
Our semi-supervised anomaly detection method allows to inspect large amounts of data for anomalies across various applications.

\end{abstract}

\section{Introduction}
Anomaly detection is the task of identifying data points that differ in their behavior compared to the majority of samples.
Reliable anomaly detection is of great interest in many real-life scenarios, especially in the context of security-sensitive systems.
Here, anomalies can indicate attacks on the infrastructure, fraudulent behavior, or general points of interest.
In recent years, the number of machine learning (ML) applications using deep learning (DL) concepts has steadily grown.
DL methods allow to analyze highly complex data for patterns that are useful to minimize a certain loss function.
Anomaly detection tasks are especially challenging for DL methods due to the inherent class imbalance.
In research, anomaly detection is often only seen as unsupervised task, thus ignoring the information gain when anomaly-related samples are available.
In our work, we develop a new DL-based anomaly detection method showing superior results with only a handful of anomaly examples -- and motivate that the very same method also works without any anomaly examples at all.

In DL-based anomaly detection, a popular idea is to use an autoencoder (AE) to preprocess, or reconstruct the input.
This type of neural network (NN) generates an output that is close to the given input under the constraint of small hidden dimensions.
Intuitively, when trained on normal samples only, the AE will miss important features that distinguish anomalous samples, thus increasing the reconstruction error.
Clearly, this method assumes that the overall error is large -- however, anomalies may be too subtle to be detected based on the output only.
In our paper, we consider the entire system context by analyzing more subtle patterns.
We show that the hidden activations of autoencoders, but also other types of NNs, are useful to judge if the current input is normal, or anomalous.
By combining the information of three interrelated NNs, we achieve strong detection results even in semi-supervised settings.

During the conceptual phase, we were inspired by coverage-guided neural network testing methods.
In this new and promising research direction, software testing concepts are transferred to DL models.
The goal is to identify faulty regions in NNs responsible for unusual behavior, or errors during run-time.
Pei et al. \cite{DBLP:journals/corr/PeiCYJ17} first introduced the idea of neuron coverage to guide a testing process.
Since then, further improvements and modifications have been proposed, e.g., by Ma et al. \cite{Ma2018} and Sun et al. \cite{Sun2018}.
Recently, Sperl et al. \cite{Sperl2019} used this concept to detect adversarial examples fed to NNs.
The authors analyze the activation values while processing benign and adversarial inputs -- a second network classifies if the recorded patterns resemble normal behavior or an attack.
In this paper, we profit from this insight and further generalize the concept by adapting it to the constraints of anomaly detection.
Whereas samples for benign as well as adversarial inputs are plentiful in adversarial ML, anomaly detection is a semi-supervised setting with only a few anomaly-related labels available.
However, also here we assume that NNs behave in a special and distinguishable manner, when confronted with anomalous data. 
We show, this behavior is detectable by analyzing the activation values during run-time.
When observing the neuron activations of NNs while processing normal inputs and synthetic anomalies, we train another NN to distinguish the nature of the analyzed data points.
Our analysis shows that even artificial anomalies train an anomaly detection model, that performs well across multiple domains, and generalizes to yet unseen anomalies.

Applying our concept, we empirically show that anomalous samples cause different hidden activations compared to normal ones.
We analyze the hidden layers of a so-called \textit{target network} by an auxiliary network, called the \textit{alarm network}.
With our \textit{anomaly network}, we automatically generate samples used during the training of the alarm network to distinguish between normal and anomalous samples.
Our evaluation shows strong results on common data sets, and we report superior performance to common baseline methods.
In summary, we make the following contributions:
\begin{itemize} 
    \item
    We propose a purely data-driven semi-supervised anomaly detection method based on the analysis of the hidden activations of NNs which we call \emph{A\textsuperscript{3}}.
    \item
    Based on our thoroughly selected set of experiments across five different data sets, we show that these patterns generalize to new anomaly types even with only a few anomaly examples available.
    \item
    We motivate that our method works in settings where no anomaly samples are available when using a generative model as anomaly network.
\end{itemize}

\section{Background and Related Work} \label{chap:preliminaries_related_work}
\subsection{Nomenclature} \label{sec:nomenclature}
Neural networks approximate an input-output mapping $f(\vect{x}; \vectg{\theta}) = \hat{\vect{y}}$ based on the learned parameters $\vectg{\theta}$.
The overall function $f=f_L \circ \ldots \circ f_1$ consists of multiple layers $f_i$.
For easier readability, we summarize the input-output relation as $f: \vect{x} \mapsto \hat{\vect{y}}$.
Generally, we denote the output, or \textit{activation}, of the $i$\textsuperscript{th} layer as $\vect{h}_{i}=\sigma(\mat{W}_{i, i-1} \cdot \vect{h}_{i-1} + \vect{b}_i)$.
Here $\sigma(\cdot)$ is a non-linear activation function, $\mat{W}_{i, j}$ and $\vect{b}_i$ the mapping parameters learned in layer $i$ with respect to layer $j$.
The input corresponds to $\vect{x} = \vect{h}_0$.
Thus, the activations simply become a function $\vect{h}_{i} (\vect{x}; \vectg{\theta})$ dependent on the input and the network parameters up to the respective layer.

The weights $\vectg{\theta}_i$ are determined by an optimization algorithm minimizing the expected loss between the desired and estimated output, $\mathcal{L}(\vect{y}, \hat{\vect{y}})$.
Our entire data set $\set{D} = \{(\vect{x}_k, \vect{y}_k)\}, \vect{x}_k \in \set{X}, \vect{y}_k \in \set{Y}$ is split into three parts: training, validation, and test.
The network weights are adapted to the training set while evaluating the performance on the validation set.
We consider categorical data, $\vect{y} \in \{1, 2, \ldots \abs{\set{Y}}\}$, where $\set{Y}$ denotes the set of available labels.
Among other things, we evaluate the transferability of A\textsuperscript{3}, i.e., the performance of the model evaluated on more labels than it was trained on, $\set{Y}_\text{train} \subset \set{Y}_\text{test}$.

\subsection{Related Work} \label{sec:related_work}
Anomaly detection is a topic of active research with a diversity of use-cases and applied methods.
For instance, in network intrusion detection \cite{Mirsky2018}, power grids \cite{Shekari2019}, or industrial control systems \cite{Feng2019}, automated mechanisms can improve the security of the overall system.
A good overview on anomaly detection in general and deep learning based systems in particular is given in the surveys \cite{Chandola2009} and \cite{Chalapathy2019}, respectively.
Note that the term ``semi-supervised'' is often ambiguous in anomaly detection.
We follow the surveys' notation, thus calling any knowledge about the underlying labels semi-supervised, e.g., also when the training data is assumed to be normal.
Famous unsupervised methods include OC-SVMs \cite{Scholkopf2000}, or Isolation Forests \cite{Liu2008a}.

In DL-based anomaly detection systems, a popular choice are architectures incorporating an AE, often used as feature extractor.
AEs are combined with classical machine learning classifiers like k-nearest neighbor \cite{Yousefi2017},
OC-SVMs for anomaly detection \cite{Andrews2016}, NN-based classifiers \cite{Qureshi2019}, or Gaussian Mixture Models \cite{Zong2018}.
Similarly, other feature extraction networks like recurrent NNs have been evaluated \cite{Nguyen2019DIoT:IoT, Malhotra2016, Schulze2019}.
To detect anomalies, AEs may also be used in their purest form: to restore the input under the constraint of small hidden layers, similar to classical dimensionality reduction methods like PCA.
The reconstruction error of samples can then be used to discriminate between normal and anomalous data points \cite{Yousefi2017, Zhou2017}.
Research has further analyzed how to improve the anomaly detection results, e.g., by iteratively adding human feedback \cite{Veeramachaneni2016, Das2017a}.

Over the past decades, computing power and data storage have steadily risen.
DL methods profit from the increased amount of training samples.
Recent research \cite{Pang2019, Ruff2020} has studied ways to incorporate known anomalies into DL-based anomaly detection.
They show that even a few anomaly labels, which are usually available in practice, improve the overall detection performance.
In our work, we extend this research area by showing that the activations of NNs differ for normal and anomalous samples.
Using our method, we are able to detect known and even new, yet unseen types of anomalies in different scenarios with high confidence. 
With our framework, we significantly improve NN-based semi-supervised anomaly detection systems.

\section{Methodology} \label{chap:architecture}
Based on the activations of a neural network $f$, we present the following hypothesis building the foundation of our paper:
\begin{displayquote}
\label{assumption}
Evaluating the activations $\vect{h}_i$ of a neural network trained on the data set $\set{D}_\text{train}$, we observe special patterns that allow to distinguish between classes the network has been trained on, and unknown classes $\vect{y}_i \notin \set{Y}_\text{train}$.
\end{displayquote}
We argue that this setting is analog to anomaly detection.
An anomaly is defined as a sample different to normal data in some unspecified behavior.
When $\set{D}_\text{train}$ describes the normal data, then any point of a yet unknown class $\vect{y}_i \notin \set{Y}_\text{train}$ defines an anomaly, i.e., a sample that does not belong to a normal class.

\subsection{Architecture}
Our goal is to map the input to a binary output;
either the sample is normal (label 0) or anomalous (label 1).
We achieve this with our new anomaly detection method comprising three parts.
Whereas the target, and alarm network are closely related to the adversarial example detection method of Sperl et al. \cite{Sperl2019}, we additionally add an anomaly network.
\begin{enumerate}
    \item
    The \textit{target network} performs a task unrelated to anomaly detection.
    In accordance to our assumption, the classes the target was trained on are considered normal.
    Several architectures of the target are possible -- we evaluate fully-connected as well as convolutional autoencoders and classifiers.
    
    \item
    The \textit{anomaly network} generates counterexamples based on the inputs, used to train the alarm network.
    We show that even a random number generator as anomaly network gives state-of-the art results.
    Furthermore, we motivate that a generative model eliminates the need for anomaly-related labels.

    \item
    The \textit{alarm network} evaluates if the given sample is normal or anomalous by observing the hidden activations of the target network.
    While training, the activations of the inputs as well as synthetic anomalies from the anomaly network are considered.
\end{enumerate}

\begin{figure}[tb]
  \begin{center}
    \includegraphics[]{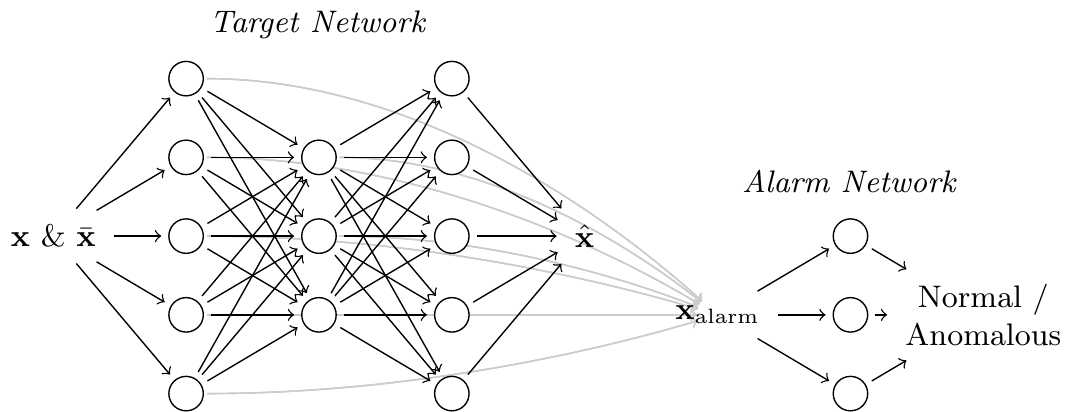}
    \caption{
    A\textsuperscript{3} consists of three parts: 1) a target network unrelated to anomaly detection (e.g., an autoencoder), 2) the anomaly network providing anomalous counterexamples $\bar{\vect{x}}$, and 3) the alarm network judging if the input $\vect{x}$ is normal.
    The three parts are connected to one overall network.
    }
    \label{fig:architecture}
  \end{center}
\end{figure}

Both parts are combined to one connected architecture.
In the scope of this paper, we fix the target network to its pretrained state.
We consider its activations caused by the input, $\vect{h}_{\text{target}, i}(\vect{x}; \vectg{\theta}_{\text{target}})$.
Our assumption is that these activations show particular patterns for samples the target network was trained on (i.e., normal samples), and for other samples (i.e., anomalous samples).
The alarm network then finds anomaly-related patterns in the target network's activations.
A high level overview is given in \Cref{fig:architecture}.

\begin{figure*}[tb]
  \begin{center}
    \includegraphics[]{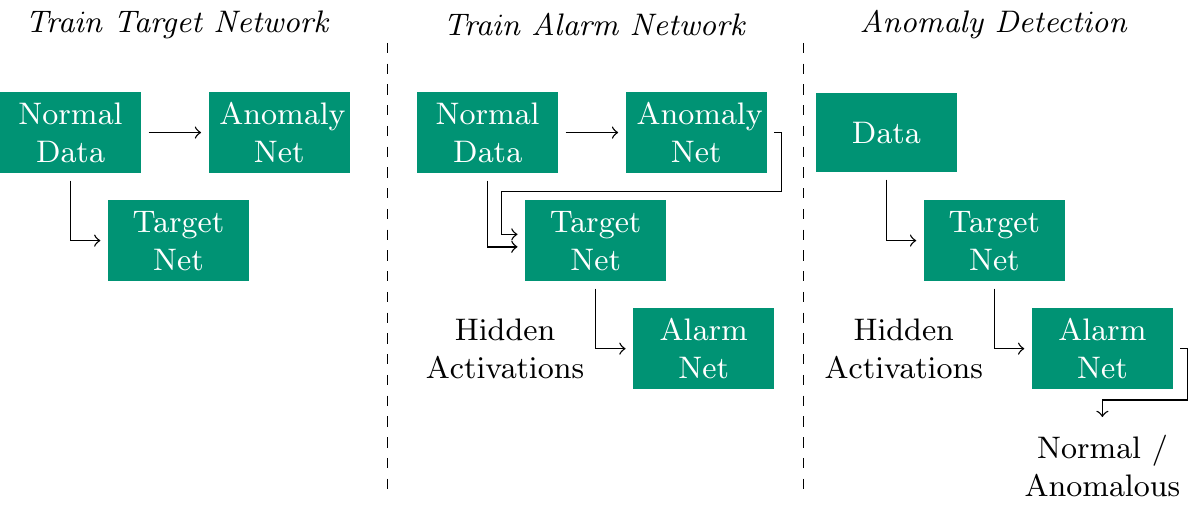}
    \caption{
    We train the target network on the training data deemed as normal. 
    To train the alarm network, the target's hidden activations caused by the input as well as the anomaly network's output are used.
    For new data points, the target's activations are evaluated by the alarm network.
    }
    \label{fig:overview}
  \end{center}
\end{figure*}

\subsection{Training Phase}
\label{sec:training_phase}
\subsubsection{Target Network}
The target network performs a task unrelated to anomaly detection on the input $\vect{x}$.
We evaluate autoencoders, $f_\text{target}: \vect{x} \mapsto \hat{\vect{x}}$, and classifiers, $f_\text{target}: \vect{x} \mapsto \hat{y} \in \{1, \ldots, \abs{\set{Y}}\}$.
According to our fundamental assumption, all samples used in the training are considered normal.

\subsubsection{Anomaly Network}
The anomaly network transforms the inputs to an anomaly, not resembling the normal data, $f_\text{anomaly}: \vect{x} \mapsto \bar{\vect{x}}$.
Thus, in accordance with the semi-supervised setting, no information about the distribution of anomalous samples is necessary.
The here generated samples are fed in the target network.
In our evaluation, we transform the input to a random realization of a normal distribution of the same dimension, i.e., $\bar{\vect{x}} \sim \mathcal{N}(\mu, \sigma^2)$.
As further outlook, we use a variational autoencoder (VAE) \cite{Kingma2014a}, i.e., an autoencoder encoding the input to the inner states $\vect{h}_{\mu}$ and $\vect{h}_{\sigma}$ forming Gaussian posteriors.
Here, anomaly samples are generated by adding Gaussian noise $\set{N}(0, 5)$ to $\vect{h}_{\mu}$, i.e., by generating highly improbable samples based on the learned distribution.

\subsubsection{Alarm Network}
The alarm network maps the input to its anomaly probability.
However, it does not operate on the input directly, but observes the target network's activations caused by the input:
\begin{displaymath}
    f_\text{alarm}:
    [\vect{h}_{\text{target}, 1}(\vect{x}; \vectg{\theta}_{\text{target}}), \ldots, \vect{h}_{\text{target}, L-1}(\vect{x}; \vectg{\theta}_{\text{target}})]
    \mapsto \hat{y} \in [0, 1].
\end{displaymath}
Hence, the output is implicitly dependent on the input and the network weights of the target and alarm model, $\hat{y}(\vect{x}; \vectg{\theta}_{\text{target}}, \vectg{\theta}_{\text{alarm}})$.
Thanks to the combination of the target, anomaly, and alarm network A\textsuperscript{3} even works when there are only a few anomalous samples available.
Further we motivate, with a suitable generative anomaly model, no anomaly samples at all is necessary.

\paragraph{Loss Function}
The alarm network is optimized on predicting the respective training labels, and classifying the anomaly network's output $\bar{\vect{x}}$ as anomalous.
We fixed the weight parameter between these objectives to $\lambda = 1.0$.
Let $\set{L}_{x}(y, \hat{y})$ denote the binary cross-entropy, the overall loss becomes:
\begin{displaymath}
    \set{L}(y, \hat{y}) = \set{L}_{x}(y, \hat{y}(\vect{x}; \vectg{\theta}_{\text{target}}, \vectg{\theta}_{\text{alarm}})) + \lambda \cdot \set{L}_{x}(1, \hat{y}(\bar{\vect{x}}; \vectg{\theta}_{\text{target}}, \vectg{\theta}_{\text{alarm}})).
\end{displaymath}

\subsection{Prediction Phase}
During the prediction phase, the target and the alarm network act as one combined system, mapping the input to a confidence interval: $f_\text{detect}: \vect{x} \mapsto \hat{y} \in [0, 1]$.
The input is transformed by the very same pipeline:
$\vect{x}$ gives rise to particular activations in the target, $\vect{h}_{\text{target}, i}(\vect{x}; \vectg{\theta}_{\text{target}})$.
Based on the target's activations, the alarm network decides whether the input is more likely to be normal, or anomalous.
\Cref{fig:overview} gives an overview about the training and prediction phase.

\section{Experiments} \label{chap:experiments}
\subsection{Data Sets}\label{sec:data_sets}
To evaluate the performance of our anomaly detection system we chose publicly available and commonly used data sets allowing the comparison to related work. 
Furthermore, we considered real world data sets and thus evaluated the applicability in complex scenarios.
We chose the following five data sets:
\begin{enumerate}
    \item
    \textbf{MNIST \cite{Lecun1998a}:}
    common image data set for ML problems with \num{70000} images showing ten handwritten digits.
    \item
    \textbf{EMNIST \cite{Cohen2017}:}
    extension to MNIST with handwritten letters.
    \item
    \textbf{NSL-KDD \cite{Tavallaee2009}:}
    common data set with around \num{150000} samples. 
    We use \textit{$\text{KDDTest}^+$} for testing containing new anomalies unseen in the training set.
    \item
    \textbf{Credit Card Transactions \cite{7376606}:}
    anonymized credit card data of around \num{285000} transactions of which \num{492} being fraudulent.
    \item
    \textbf{CSE-CIC-IDS2018 \cite{Sharafaldin2018}:}
    large network data set.
    We omit the DDoS data due to the high resource demands.
    Around five million samples remain.
\end{enumerate}
We limit the preprocessing to minmax-scaling numerical and 1-Hot-encoding categorical data.
Samples still containing non-numerical values are omitted.
Generally, we took 80\% of the data for training, 5\% for validation, and 15\% for testing.
If a test set is given, we used it instead.

\subsection{Baseline Methods}
We compared the performance of A\textsuperscript{3} to four common baseline methods.
Note that we only considered baseline methods that scale to the large amount of data.

\paragraph{Autoencoder Reconstruction Threshold}
When trained on normal data only, there is a measurable difference in the reconstruction error of autoencoders when an anomalous sample is fed in.
We calculate the mean squared error, i.e., $\set{L}(\vect{x}, \hat{\vect{x}}) = \norm{\hat{\vect{x}} - \vect{x}}^2_2$, to quantify this difference.
As some target networks use an autoencoder architecture, we take the very same models for this baseline.

\paragraph{Isolation Forest (IF)}
IF is a commonly used unsupervised anomaly detection method by Liu et al. \cite{Liu2008a}.
Based on the given data, an ensemble of random trees is built.
The average path length results in an anomaly score.
We use the implementation provided by scikit-learn \cite{scikit-learn} along with the default parameters.

\paragraph{Deep Autoencoding Gaussian Mixture Model (DAGMM)}
DAGMM is a state-of-the-art unsupervised DL-based anomaly detection method by Zong et al. \cite{Zong2018}.
The authors combine information of an autoencoder with a Gaussian mixture model.
In accordance with their first set of experiments, we take normal samples for the training.
For all experiments, we use the implementation by Nakae \cite{Nakae}, and the architecture recommended for the KDDCUP data set.

\paragraph{Deviation Networks (DevNet)}
DevNet is a state-of-the-art semi-supervised DL-based anomaly detection method by Pang et al. \cite{Pang2019}.
The anomaly detection is split between a feature learner and an anomaly score learner, both implemented as an NN.
The anomaly scorer learns to score normal samples close to a Gaussian prior distribution, and enforces a minimum distance to anomalous samples.
As recommended by the authors, we use their default architecture.

\subsection{Anomaly Detection Constraints}\label{sec:anomaly_detection_constraints}
Special constraints apply in the setting of anomaly detection.
With our experiments, we show that A\textsuperscript{3} performs well nonetheless.

\begin{enumerate}
    \item \textit{Scarcity of anomaly samples.}\label{constraint1}
    Generally, many samples are required for strong performance using DL frameworks.
    However, there is a natural imbalance in anomaly detection scenarios:
    most data samples are normal, only a few examples of anomalies were found manually.
    We show that our anomaly detection algorithm performs well in this semi-supervised setting.
    
    \item \textit{Variable extend of abnormality.}\label{constraint2}
    Anomalous samples are not bound by a common behavior or magnitude of abnormality.
    By definition, the only difference is that anomalies do not resemble normal samples.
    We show the \textit{transferability} of our algorithm, i.e., known anomalies during training also reveal yet unknown anomalies during testing.

    \item \textit{Driven by data, not expert knowledge.}\label{constraint3}
    A suitable anomaly detection algorithm should be applicable to multiple settings, even when no expert knowledge is available.
    Performance that is only achievable using domain knowledge, may result in inferior results in other settings.
    We show that our algorithm \textit{generalizes} to other settings, i.e., uses the data itself to distinguish between normal and anomalous behavior.
\end{enumerate}

\begin{table}[tb]
\caption{Experiments exploring the detection of known \& unknown anomalies.}
\label{exp:all}
\begin{center}
\begin{tabular}{c c c c c c}
\toprule
 & \textbf{Data}   & \textbf{Normal}   & \textbf{Train Anomaly}    & $\subseteq$ & \textbf{Test Anomaly}    \\
\midrule
1a\&4a & MNIST               & 0, \ldots, 5      & 6, 7                      & & 6, 7 \\
1b\&4c & MNIST               & 4, \ldots, 9      & 0, 1                      & & 0, 1 \\
1c & NSL-KDD             & Normal            & \makecell{DoS, Probe}      & & \makecell{DoS, Probe}\\
1d & NSL-KDD             & Normal            & \makecell{R2L, U2R}      & & \makecell{R2L, U2R}\\
1e & IDS                 & Benign            & \makecell{BF, Web, DoS, Infil.}            & & \makecell{BF, Web, DoS, Infil.} \\
1f & IDS                 & Benign            & \makecell{Bot, Infil., Web, DoS}            & & \makecell{Bot, Infil., Web, DoS} \\
1g & CC             & Normal            & \makecell{Fraudulent}      & & \makecell{Fraudulent}\\
\midrule
2a\&4b & MNIST               & 0, \ldots, 5      & 6, 7                      & & 6, 7, 8, 9 \\
2b\&4d & MNIST               & 4, \ldots, 9      & 0, 1                      & & 0, 1, 2, 3 \\
2c & NSL-KDD             & Normal            & \makecell{DoS, Probe}      & & \makecell{DoS, Probe, R2L, U2R}\\
2d & NSL-KDD             & Normal            & \makecell{R2L, U2R}      & & \makecell{R2L, U2R, DoS, Probe}\\
2e & IDS                 & Benign            & \makecell{BF, Web, DoS, Infil.}            & & \makecell{BF, Web, DoS, Infil., Bot} \\
2f & IDS                 & Benign            & \makecell{Bot, Infil., Web, DoS}            & & \makecell{Bot, Infil., Web, DoS, BF} \\
\midrule
3a & \makecell{(E-)MNIST} & 0, \ldots, 9      & A, B, C, D, E          & & \makecell{A, B, C, D, E} \\
3b & \makecell{(E-)MNIST} & 0, \ldots, 9      & A, B, C, D, E          & & \makecell{A, B, C, D, E, V, W, X, Y, Z} \\
3c & \makecell{(E-)MNIST} & 0, \ldots, 9      & V, W, X, Y, Z          & & \makecell{V, W, X, Y, Z} \\
3d & \makecell{(E-)MNIST} & 0, \ldots, 9      & V, W, X, Y, Z          & & \makecell{A, B, C, D, E, V, W, X, Y, Z} \\
\bottomrule
\end{tabular}
\end{center}
\end{table}

\subsection{Experimental Setup} \label{sec:experimental_setup}
We designed multiple experiments to show that A\textsuperscript{3} works under all three constraints.
An overview is given in \Cref{exp:all}.

\paragraph{Experiment 1: Detection of Known Anomalies}
\label{experiment1}
Considering constraint~\ref{constraint1}, we evaluated the fundamental assumption of our method, i.e.:
the activation values of the target network contain information to distinguish between normal and anomalous samples.
It is important to remember that only the alarm network, not the target, is trained on the anomaly detection task.
We limited the train anomaly samples to \{5, 25, 50, 100\} randomly selected instances in accordance with the semi-supervised setting.
Note, this limitation may cause some classes not to be present during training.

\paragraph{Experiment 2: Transferability to Unknown Anomalies}
\label{experiment2}
Considering constraints~\ref{constraint1} and \ref{constraint2}, we evaluated the transferability of our fundamental assumption, i.e.:
the activation values of the target network are inherently different for normal and anomalous samples.
Similar to experiment~1, we evaluated the detection performance bound by the scarcity of anomaly labels.
Furthermore, the test data set contained more anomaly classes than the alarm network has been trained on.
In other words, we tried to find anomalies that follow a different nature and data distribution than the one of the few known samples.

\paragraph{Experiment 3: Generality of the Method}
\label{experiment3}
Considering constraints~\ref{constraint1},~\ref{constraint2},~and~\ref{constraint3}, we evaluated the generality of our fundamental assumption, i.e.:
the activation values of any type of target network contain information to distinguish between normal and anomalous samples.
We used a publicly available classifier as target, extracted the activation values, and tested whether these can be used to detect known as well as unknown anomalies.
Hence, we motivate that our anomaly detection mechanism can be applied to already existing target networks and environments of any type.

\paragraph{Experiment 4: Outlook to Unsupervised Anomaly Detection}
\label{experiment4}
Considering the extreme case of constraint~\ref{constraint1}, we made first evaluations of the detection performance when no labeled anomalies are available during training.
We solely used normal samples, as well as the output of a generative anomaly network to train A\textsuperscript{3}. 

\begin{table}[htb]
\caption{Dimensionality of the layers. All layers are activated by ReLUs.}
\label{table:architectures}
\begin{center}
\begin{tabular}{c c c c c}
\toprule
\textbf{Data Set}   & \textbf{Target Architecture}   & \textbf{Alarm Architecture} \\
\midrule
MNIST               & according to \cite{Chollet2016} & 1000, 500, 200, 75, 1  \\
NSL-KDD   & 200, 100, 50, 25, 50, 100, 200                     & 1000, 500, 200, 75, 1  \\
IDS       & 150, 80, 40, 20, 40, 80, 150 & 1000, 500, 200, 75, 1    \\
CreditCard       & 50, 25, 10, 5, 10, 25, 50 & 1000, 500, 200, 75, 1    \\
\makecell{MNIST \& EMNIST} & according to \cite{CholletCNN} & 1000, 500, 200, 75, 1   \\
\bottomrule
\end{tabular}
\end{center}
\end{table}

\subsection{Experiment Overview}
An overview about the used architectures for each experiment is found in~\Cref{table:architectures}.
For the symmetric AE-based target models, we chose the first layer to be slightly larger than the dimension of the input vectors, whereas the hidden representation should be smaller.
For the sake of simplicity, we used a common alarm model architecture throughout this paper.
Note that for the MNIST-related experiments, we considered two publicly available architectures from Keras \cite{chollet2015keras}, i.e., a convolutional AE \cite{Chollet2016} extended by a dropout layer for experiment~1~and 4, as well as a CNN \cite{CholletCNN} for experiment~3.
This underlines the generality of our method.
All layers are activated by ReLUs.

\paragraph{Parameter Choices}
Based on a non-exhaustive parameter search on MNIST, we chose the following global optimizer settings:
Adam \cite{Kingma2014} with a learning rate of 0.001 for the target network, and 0.00001 for the alarm network was used.
The training was stopped after 30 and 60 epochs, respectively.
No other regularizer than 10\% dropout \cite{Hinton2014} before the last layer was used.
We ran our experiments on an Intel Xeon E5-2640 v4 server accelerated by an NVIDIA Titan X GPU.

\paragraph{Anomaly Network}
For experiments~1,~2,~\&~3, we used a simple Gaussian noise generator as anomaly network.
As all inputs are within $[0,1]$, we fixed the noise parameters to $\set{N}(.5, 1)$.
For experiment~4, we chose a VAE with the dense hidden layers $800, 400, 100, 25, 100, 400, 800$.
During the training of the target network, it was adapted to reconstruct the normal samples.

\begin{figure}[htb]
  \begin{center}
    \includegraphics[]{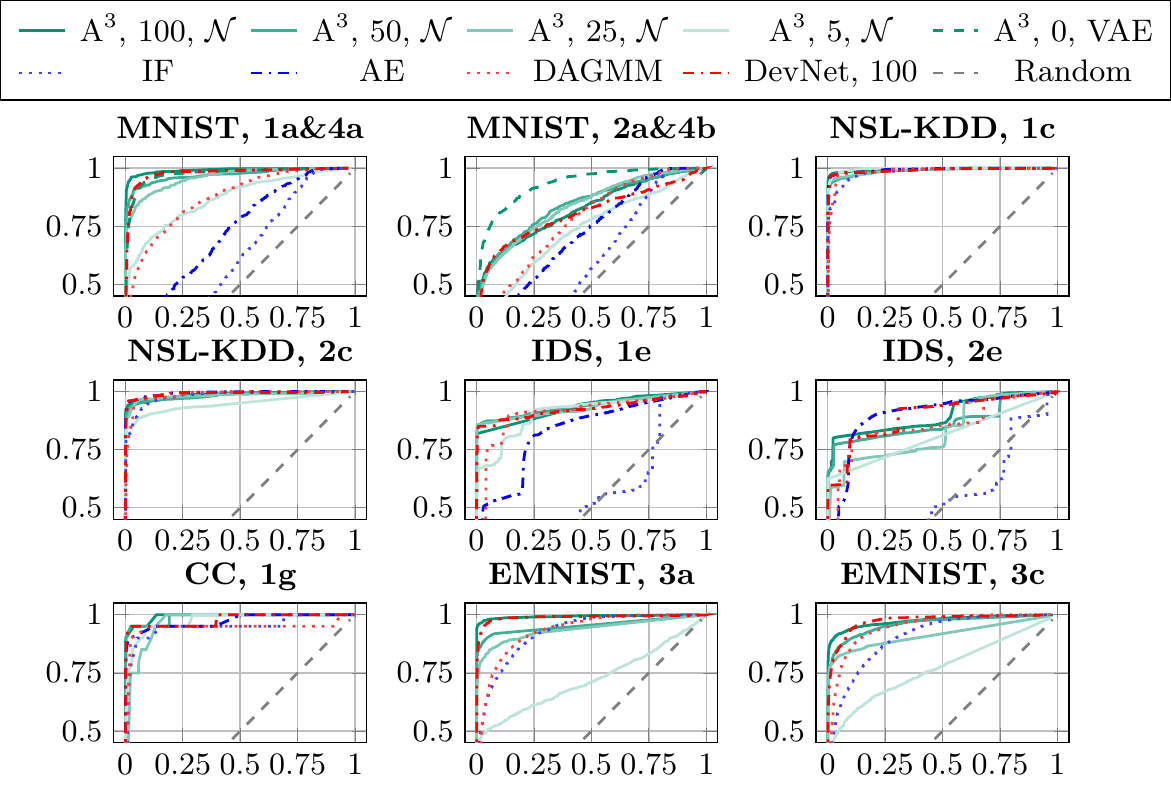}
    \caption{
    ROC curves showing \emph{the true positive rate vs. the false positive rate} evaluated on the validation set for several experiments.
    For A\textsuperscript{3}, we capped the amount of anomaly samples during training.
    We used a noise generator ($\set{N}$) as anomaly network, and for experiment~4 a VAE.
    }
    \label{fig:roc}
  \end{center}
\end{figure}

\section{Results and Evaluation} \label{chap:results}
In the following, we present and evaluate the results measured during our experiments.
We show the receiver operating characteristic (ROC) curve, i.e., the true positive rate (TPR) as a function of the false positive rate (FPR), evaluated on the validation sets.
For the final results, evaluated on the test sets, we use the average precision (AP), and the area under the ROC curve (AUC) as metrics to be consistent with the related work \cite{Pang2019}.
Whereas the AP quantifies the trade-off between precision and recall, the AUC measures the trade-off between the TPR and FPR.
Both metrics are independent of the chosen detection threshold, and thus give a good overview about the general detection performance.

\subsection{Validation Results}
In \Cref{fig:roc}, we show the ROC curves of the first experiments for each validation data set.
To simulate real-life conditions, where usually just a few anomaly samples are available, we restricted the known anomalies to \{5, 25, 50, 100\} for A\textsuperscript{3}. 
Note, these anomalies are sampled randomly from all available training anomaly samples -- it could well be that they are restricted to one class.
We report that A\textsuperscript{3} generally follows an intuitive behavior: the more known anomalies are used, the better the anomaly detection performance.
Even for 5 samples an adequate performance is possible on most data sets.
As little as 25 samples are needed to surpass the performance of the state-of-the-art unsupervised anomaly detection methods.

Moreover, A\textsuperscript{3} surpasses, or matches the performances of the state-of-the-art semi-supervised anomaly detection method.
An important feature is the steep rise of the TPR, thus less false positives are seen resulting in a strong detection performance.
To choose a suitable detection threshold in practice, a very low FPR has to be tolerated using our method.
This significantly reduces manual work, and builds trust in the detection results.

\begin{table}[tb]
\caption{Test results given all normal, and 100 anomaly samples.}
\label{table:final_results}
\begin{center}
\begin{tabular}{c  c c | c c | c c | c c | c c}
\toprule
& \multicolumn{2}{c}{\textbf{A\textsuperscript{3}}} & \multicolumn{2}{c}{\textbf{AE}} & \multicolumn{2}{c}{\textbf{IF}} & \multicolumn{2}{c}{\textbf{DAGMM}} & \multicolumn{2}{c}{\textbf{DevNet}}\\
    & AP & AUC  &  AP & AUC  &  AP & AUC  &  AP & AUC  &  AP & AUC\\
\midrule
1a & \underline{.98}$\pm$.00 & \underline{.99}$\pm$.00 & .44$\pm$.04 & .69$\pm$.03 & .27$\pm$.02 & .56$\pm$.02 & .71$\pm$.01 & .85$\pm$.01 & .96$\pm$.01 & .98$\pm$.00\\
1b & \underline{.99}$\pm$.00 & \underline{1.0}$\pm$.00 & .25$\pm$.03 & .40$\pm$.04 & .42$\pm$.02 & .53$\pm$.01 & .34$\pm$.03 & .64$\pm$.04 & .98$\pm$.01 & .99$\pm$.01\\
1c & .96$\pm$.01 & .96$\pm$.01 & .96$\pm$.01 & .96$\pm$.00 & \underline{.98}$\pm$.00 & \underline{.97}$\pm$.00 & .90$\pm$.04 & .94$\pm$.01 & .95$\pm$.01 & .96$\pm$.01\\
1d & .76$\pm$.04 & .88$\pm$.05 & .51$\pm$.04 & .81$\pm$.02 & .48$\pm$.02 & .84$\pm$.00 & .59$\pm$.03 & \underline{.90}$\pm$.01 & \underline{.78}$\pm$.03 & .88$\pm$.04\\
1e & \underline{.89}$\pm$.03 & \underline{.94}$\pm$.01 & .73$\pm$.12 & .83$\pm$.14 & .17$\pm$.01 & .46$\pm$.03 & .63$\pm$.13 & .87$\pm$.08 & .89$\pm$.01 & .93$\pm$.00\\
1f & \underline{.90}$\pm$.01 & \underline{.94}$\pm$.01 & .57$\pm$.11 & .84$\pm$.06 & .13$\pm$.00 & .35$\pm$.03 & .43$\pm$.16 & .73$\pm$.13 & .73$\pm$.04 & .90$\pm$.00\\
1g & \underline{.78}$\pm$.09 & .96$\pm$.01 & .52$\pm$.09 & .97$\pm$.00 & .15$\pm$.04 & .96$\pm$.01 & .38$\pm$.25 & .96$\pm$.02 & .76$\pm$.04 & \underline{.98}$\pm$.01\\
\midrule
2a & \underline{.87}$\pm$.03 & \underline{.88}$\pm$.03 & .63$\pm$.02 & .72$\pm$.02 & .40$\pm$.01 & .54$\pm$.02 & .69$\pm$.01 & .75$\pm$.02 & .82$\pm$.02 & .82$\pm$.03\\
2b & \underline{.92}$\pm$.02 & \underline{.92}$\pm$.02 & .61$\pm$.01 & .63$\pm$.01 & .61$\pm$.01 & .64$\pm$.01 & .61$\pm$.01 & .72$\pm$.02 & .90$\pm$.03 & .90$\pm$.04\\
2c & .94$\pm$.01 & .92$\pm$.02 & .94$\pm$.01 & .93$\pm$.01 & \underline{.96}$\pm$.00 & \underline{.94}$\pm$.00 & .91$\pm$.03 & .93$\pm$.01 & .94$\pm$.00 & .92$\pm$.01\\
2d & .94$\pm$.03 & .92$\pm$.03 & .94$\pm$.01 & .93$\pm$.01 & \underline{.96}$\pm$.00 & \underline{.94}$\pm$.00 & .91$\pm$.03 & .93$\pm$.01 & .89$\pm$.02 & .88$\pm$.03\\
2e & \underline{.87}$\pm$.02 & .90$\pm$.03 & .80$\pm$.05 & \underline{.91}$\pm$.02 & .19$\pm$.01 & .44$\pm$.02 & .48$\pm$.12 & .73$\pm$.11 & .83$\pm$.02 & .90$\pm$.01\\
2f & \underline{.90}$\pm$.03 & \underline{.93}$\pm$.02 & .81$\pm$.07 & .90$\pm$.03 & .19$\pm$.01 & .44$\pm$.02 & .57$\pm$.06 & .81$\pm$.05 & .82$\pm$.02 & .92$\pm$.00\\
\midrule
3a & .98$\pm$.01 & \underline{.99}$\pm$.01 & - & - & .85$\pm$.01 & .93$\pm$.00 & .83$\pm$.03 & .93$\pm$.01 & \underline{.99}$\pm$.01 & .99$\pm$.00\\
3b & .97$\pm$.01 & .96$\pm$.01 & - & - & .89$\pm$.01 & .91$\pm$.01 & .93$\pm$.01 & .95$\pm$.00 & \underline{.98}$\pm$.00 & \underline{.98}$\pm$.00\\
3c & \underline{.99}$\pm$.00 & \underline{.99}$\pm$.00 & - & - & .79$\pm$.02 & .89$\pm$.01 & .90$\pm$.01 & .96$\pm$.00 & .96$\pm$.01 & .98$\pm$.01\\
3d & .95$\pm$.02 & \underline{.95}$\pm$.02 & - & - & .89$\pm$.01 & .91$\pm$.01 & .93$\pm$.01 & .95$\pm$.00 & \underline{.95}$\pm$.03 & .94$\pm$.04\\
\bottomrule
\end{tabular}
\end{center}
\end{table}

\subsection{Test Results}
In \Cref{table:final_results}, we summarize the results averaged over four passes using the test data sets.
To simulate real-life conditions, we limit the amount of available anomalies to 100 randomly chosen samples of the training set.
We report that A\textsuperscript{3} performs well even under this strict setting.
It well surpasses the unsupervised baseline methods, and on most experiments also the semi-supervised baseline method.
The performance remains strong across all data sets and experiments.
We conclude that our main fundamental assumption, i.e., the hidden activations of NNs carry information useful to anomaly detection, is well supported.

Moreover, we see strong evidence for the hypotheses made throughout the experiments.
In experiment~1, we tried to find known anomalies.
The test results show the highest results for this setting, thus our method very well identifies suitable patterns for this task.
These patterns generalize well to yet unseen patterns as shown in experiment~2.
Although, only parts of the test anomalies are known during training, strong results are achieved.
Whereas the aforementioned experiments used deep \& convolution autoencoders as target network, we generalize the setting to deep classifiers in experiment~3.
Also here, superior results are achieved.
We conclude that A\textsuperscript{3} is able to detect known, and yet unknown anomalies with high confidence, and is flexible enough to adapt to a wide range of environments.

\subsection{Outlook to Unsupervised Anomaly Detection}
In the supplementary experiment~4, we further motivate that A\textsuperscript{3} performs well even when no anomaly sample is available.
For this, we chose a generative model as anomaly network, a VAE in our case.
In \Cref{fig:roc}, we see a superior performance on the validation set compared to other unsupervised anomaly detection methods.
We assume that sampling from the improbable distribution regions, which the VAE has learned, is a suitable way to generate counterexamples useful to the alarm network.
In \Cref{table:outlook_results}, we summarize our results on the test set.
Whereas, experiment~4a\&b show very promising results improving on the basic anomaly network, the variance for 4c\&d is relatively high.
We are confident that more parameter tuning will lead to consistent performance over all data sets.
Future research may leverage these results to a more general setting.
We are happy to report consistently strong results when using a simple noise generator as anomaly network shown in experiment 1,~2,~\&~3.
With this setting, we provide a ready-to-use end-to-end anomaly detection framework achieving state-of-the-art performance even in strict semi-supervised environments for a variety of use-cases. 

\begin{table}[tb]
\caption{Test result for exp.~4, where no anomaly samples were used to train A\textsuperscript{3}.}
\label{table:outlook_results}
\begin{center}
\begin{tabular}{c c @{\hskip 2em} c c @{\hskip 2em} c c @{\hskip 2em} c c }
\textbf{4a}-AP & \textbf{4a}-AUC  &  \textbf{4b}-AP & \textbf{4b}-AUC  &  \textbf{4c}-AP & \textbf{4c}-AUC  &  \textbf{4d}-AP & \textbf{4d}-AUC \\
\midrule
.95$\pm$.01 & .98$\pm$.00 & .94$\pm$.02 & .95$\pm$.02 & .55$\pm$.07 & .67$\pm$.06 & .60$\pm$.09 & .62$\pm$.10\\
\end{tabular}
\end{center}
\end{table}

\section{Discussion \& Future Work} \label{chap:discussion}
In this paper, we built an anomaly detection method based on the analysis of hidden activations.
By observing the activations of an NN that was trained on a task unrelated to anomaly detection, we take all context information it has learned into consideration.
Future research may leverage this method to other use-cases, and further formalize the explored theory.
A\textsuperscript{3} shows strong anomaly detection results for all five analyzed data sets across all experiments.
We motivate that the activation analysis generalizes to yet unseen anomalies across different network architectures.
Thanks to the modularity of our concept, various architectures may be used as target, alarm or anomaly networks covering numerous types of data and use-cases.
Future work may integrate other powerful architectures, e.g., generative adversarial networks \cite{Goodfellow2014b} as already applied to other anomaly detection settings \cite{Zenati2018, Akcay2019}.
We emphasize the real-life applicability by limiting the amount of anomaly samples during training.
In practice, often a few known anomalies are available, e.g., by manual exploration, or unsupervised methods.
Good performance was already achieved with a handful of anomaly samples.
During our research, we saw further improvements with a shifted output regularizer, e.g., $\set{L}_{\text{reg}}(y) = \lambda \cdot \abs{1 - y}$, favoring the detection of anomalies.
With this set of regularizers, and the presented outlook to generative anomaly networks, we leave developers the ability to further tune the performance of A\textsuperscript{3} in their implementation.

\section{Conclusion} \label{chap:conclusion}
We introduce a novel approach for anomaly detection called A\textsuperscript{3} based on the hidden activation patterns of NNs.
Our architecture comprises three parts: a target network unrelated to the anomaly detection task, an anomaly network generating anomalous training samples, and an alarm network analyzing the resulting activations of the target.
Our framework works under common assumptions and constraints typically found in anomaly detection tasks.
We assume that anomalous training samples are scarce, and new types of anomalies exist during deployment.
With our evaluation, we provide strong evidence that our method works on different target network architectures, and generalizes to yet unseen anomalous samples.
Furthermore, we detect anomalies across different data types with just a few or even no labeled anomalies at all available during training.
We present a valuable semi-supervised DL-based anomaly detection framework providing a purely data-driven solution for a variety of use-cases.


\bibliographystyle{splncs04}
\bibliography{references}{}

\end{document}